\documentclass[preprintnumbers,aps,prl,twocolumn,superscriptaddress,notitlepage,nofootinbib]{revtex4-1}
\usepackage{amsmath, bm}
\usepackage{epsfig}
\usepackage{color,soul}
\usepackage{multirow}
\newcommand{\bea}{\begin{eqnarray}}
\newcommand{\eea}{\end{eqnarray}}

\newcommand{\bef}{\begin{figure}[h!tb]\centering}
\newcommand{\eef}{\end{figure}}
\newcommand{\be}{\begin{equation}}
\newcommand{\ee}{\end{equation}}

\begin{document}
\title{First extraction of the $\Lambda$ polarising fragmentation function from Belle $e^+e^-$ data}

\author{Umberto D'Alesio}
\email{umberto.dalesio@ca.infn.it}
\affiliation{Dipartimento di Fisica, Universit\`a di Cagliari, Cittadella Universitaria, I-09042 Monserrato (CA), Italy}
\affiliation{INFN, Sezione di Cagliari, I-09042 Monserrato (CA), Italy}

\author{Francesco Murgia}
\email{francesco.murgia@ca.infn.it}
\affiliation{INFN, Sezione di Cagliari, I-09042 Monserrato (CA), Italy}

\author{Marco Zaccheddu}
\email{marco.zaccheddu@ca.infn.it}
\affiliation{Dipartimento di Fisica, Universit\`a di Cagliari, Cittadella Universitaria, I-09042 Monserrato (CA), Italy}
\affiliation{INFN, Sezione di Cagliari, I-09042 Monserrato (CA), Italy}

\date{\today}

\begin{abstract}
We present a thorough phenomenological analysis of the experimental data from Belle Collaboration for the transverse $\Lambda$ and $\bar\Lambda$ polarisation, measured in $e^+e^-$ annihilation processes, for the case of inclusive (plus a jet) and associated production with a light charged hadron. This allows for the \emph{first ever} extraction of the quark polarising fragmentation function for a $\Lambda$ hyperon, a transverse momentum dependent distribution giving the probability that an unpolarised quark fragments into a transversely polarised spin-1/2 hadron.
\end{abstract}

\date{\today}

\maketitle
{\it Introduction.}
The internal structure of the nucleon as well as the parton fragmentation mechanism into hadrons are key issues in hadron physics. In the last years, it has become clear that a deeper understanding of these phenomena requires a more complete and detailed view. This can be achieved by moving from a collinear picture to a three-dimensional description, namely introducing transverse momentum dependent distributions (shortly referred to as TMDs). Concerning the distribution sector, when also spin degrees of freedom are included, the most studied TMD function is certainly the Sivers function~\cite{Sivers:1989cc,Sivers:1990fh}. This gives the azimuthal asymmetry in the distribution of quarks or gluons within a fast moving transversely polarised nucleon. Many extractions are now available and its phenomenology is well consolidated. In the fragmentation sector an analogous role is played by the Collins fragmentation function (FF)~\cite{Collins:1992kk}, giving the asymmetric azimuthal distribution of an unpolarised hadron in the fragmentation of a transversely polarised quark. Also in this case we have reached a clear evidence.
A much less explored TMD is the so-called polarising fragmentation function, giving the distribution of a transversely polarised spin-1/2 hadron coming from the fragmentation of an unpolarised quark. Among the main properties of this TMD-FF, we recall that it is T-odd, like the Collins and the Sivers functions, but chiral even, like the Sivers function. This last feature allows one to access it directly without any unknown, chiral-odd, counterpart.
This TMD-FF can then give complementary and relevant information on the fragmentation mechanism, and could help in understanding the so far unknown origin of the sizeable transverse polarisation of hyperons observed in many inclusive hadronic processes~\cite{Bunce:1976yb,Heller:1978ty}.

Introduced in Ref.~\cite{Mulders:1995dh}, it was studied phenomenologically in Ref.~\cite{Anselmino:2000vs}, where the puzzling transverse polarisation data for the inclusive production of $\Lambda$ hyperons in unpolarised hadron-hadron collisions were considered. Within a simple TMD model (TMD factorisation for these single-scale processes is still missing), some of its interesting features were extracted and a quite good description of data was achieved.
For the processes under study, namely $e^+e^-\to \Lambda ({\rm jet}) + X$, $e^+e^-\to \Lambda\, h + X$, where two well separated energy scales are present (the large $Q^2$ of the virtual photon and the small relative transverse momentum between the $\Lambda$ and the jet (or $h$)), their treatment in terms of TMDs is formally proven and well consolidated~\cite{Collins:2011zzd,GarciaEchevarria:2011rb}. We notice here that the transverse $\Lambda$ polarisation in a process with these same features, namely $\ell p\to \ell' \Lambda^\uparrow + X$, was discussed by two of us in Ref.~\cite{Anselmino:2001js}.

The new available data from the Belle Collaboration at KEK~\cite{Guan:2018ckx} on transverse $\Lambda/\bar\Lambda$ hyperon polarisation in $e^+e^-$ processes have triggered a renewed interest on these phenomena. As we will show, they allow for the first ever extraction of the polarising FF in a clear and accurate way.
It is worth stressing that, since no other contribution from the initial state could play a role, this process is unique, and the cleanest one, in accessing this TMD-FF.
A preliminary phenomenological study, even though in a simplified scheme, has been discussed in Ref.~\cite{Anselmino:2019cqd}. Here we present a detailed global analysis of Belle data at the same level of accuracy as that of current studies on other relevant TMDs.

{\it Formalism.}
We consider the processes $e^+e^-\to h_1 ({\rm jet}) + X$ and $e^+e^-\to h_1 h_2 + X$, where $h_1$ is a spin-1/2 hadron and the second (light and unpolarised) hadron, $h_2$, is produced almost back-to-back with respect to $h_1$.
Adopting the helicity formalism with inclusion of transverse momentum effects one can obtain the most general expressions of all polarisation observables for these processes. The detailed calculation, with the complete classification of the quark and gluon TMD fragmentation functions for a spin-1/2 hadron, will be presented in a forthcoming paper~\cite{D'Alesio:2020xxx} and is in agreement with the results in Ref.~\cite{Boer:1997mf}. Here we focus directly on the case of the transverse polarisation for which data are now available.

This quantity is defined as
\begin{equation}
{\cal P}_T = \frac{d\sigma^{\uparrow} - d\sigma^{\downarrow} }{d\sigma^{\uparrow} + d\sigma^{\downarrow}} = \frac{d\sigma^{\uparrow} - d\sigma^{\downarrow} }{d\sigma^{\rm unp}} ,
\label{polt}
\end{equation}
where $d\sigma^{\uparrow(\downarrow)}$ is the differential cross section for the production of a hadron transversely polarised along  the up (down) direction with respect to the production plane (see below) and $d\sigma^{\rm unp}$ is the unpolarised cross section.

For inclusive production (within a jet), the polarisation is measured orthogonally to the \emph{thrust plane}, containing the jet (more precisely the thrust axis, $\hat{\bm{T}}$) and the spin-1/2 hadron momentum, $\bm{P}_{h_1}$, that is along $\hat{\bm{T}}\times\bm{P}_{h_1}$. For the associated production with a light hadron, $h_2$, one considers the \emph{hadron plane},  containing the two hadrons, and the transverse polarisation is measured along $(-\bm{P}_{h_2}\times \bm{P}_{h_1})$.

For the first case, in a leading order approach, we choose a frame so that the $e^+e^-\to q\bar q$ scattering occurs in the $\widehat{xz}$ plane, with $\theta$ being the angle between the back-to-back quark-antiquark (identifying the $\bm{z}$ axis) and the $e^+e^-$ directions.
The three-momentum of the hadron $h_1$, with mass $m_{h_1}$, light-cone momentum fraction, $z_1$, and intrinsic transverse momentum,  $\bm{p}_{\perp 1}$, with respect to the direction of the fragmenting quark, is given as
\bea
\bm{P}_{h_1} & = & \Big(p_{\perp 1}\cos \varphi_1, p_{\perp 1}\sin \varphi_1, z_1 \frac{\sqrt{s}}{2}(1-a_{h_1}^2/z_1^2)\Big)\,,
\eea
 {\rm with}
\be
a^2_{h_1} = (p_{\perp 1}^2 + m_{h_1}^2)/s\,.
\ee
The transverse polarisation is then given as~\cite{D'Alesio:2020xxx}
\bea
{\cal P}_T(z_1, p_{\perp 1}) &  = & \frac{\sum_{q} e^2_q (1 + \cos^2\theta) \Delta D_{h_1^\uparrow/q}(z_1,p_{\perp 1})}{ \sum_{q}  e^2_q (1 + \cos^2\theta) D_{h_1/q}(z_1,p_{\perp 1})}\,,
\label{PolTjet}
\eea
where the sum runs over quark and antiquarks, and $\Delta D_{h_1^\uparrow\!/q}$ is the polarising FF, also denoted as $D_{1T}^{\perp q}$. The two FFs are related as follows~\cite{Bacchetta:2004jz}:
\be
\Delta D_{h^\uparrow\!/q}(z,p_{\perp}) = \frac{p_{\perp}}{z m_{h}} D_{1T}^{\perp q}(z,p_{\perp})\,.
\ee
More precisely, for a hadron with polarisation vector $\hat{\bm{P}}\equiv \uparrow$, coming from the fragmentation of a quark with momentum $\bm{p}_q$, the polarising FF is defined as
\bea
\label{polFF}
\Delta \hat D_{h^\uparrow\!/q}(z,\bm{p}_{\perp}) &\equiv & \hat D_{h^\uparrow\!/q}(z,\bm{p}_\perp) - \hat D_{h^\downarrow\!/q}(z,\bm{p}_\perp) \nonumber\\
&=& \Delta D_{h^\uparrow\!/q}(z,p_{\perp}) \,\hat{\bm{P}} \cdot (\hat{\bm{p}}_q \times \hat{\bm{p}}_\perp) \,.
\eea
Eq.~(\ref{PolTjet}) is obtained by integrating the numerator and the denominator in Eq.~(\ref{polt}), differential in $z_1,\bm{p}_{\perp 1}, \cos\theta$, over $\varphi_1$ and refers to the transverse polarisation in the hadron helicity frame (i.e.~transverse w.r.t.~the plane containing the fragmenting quark and the hadron $h_1$). As a matter of fact, at leading order, this coincides with the thrust-plane frame adopted in the experimental analysis. For massive hadrons, two further scaling variables are usually introduced: the energy fraction $z_h = 2E_h/\sqrt s$ (adopted in Belle analysis), and the momentum fraction $z_{p} = 2 |\bm{P}_{h}|/\sqrt s$. These are related as:
\bea
z_{h,p} &\simeq & z\,\big[1 \pm m_h^2/(z^2s)\big] \label{zhp}\\
z_p & = & z_h \big [1-4\,m_h^2/(z_h^2s)\big ]^{1/2} \label{zpzh}\,,
\eea
where the expression (\ref{zhp}) is valid at ${\cal O}(p_\perp/(z\sqrt s))$, for $p_\perp\sim \Lambda_{\rm QCD}$.
We will use this approximation here and in the following, while keeping, for $\Lambda$ hyperons, the full dependence on $m_{h}$.

For the associated production, in accordance with the experimental analysis, we adopt the following configuration: the produced unpolarised hadron ($h_2$ in our case) identifies the $z$ direction [$\bm{P}_{h_2}= (0,0,-|\bm{P}_{h_2}|)$] and the $\widehat{xz}$ plane is determined by the lepton and the $h_2$ directions (with the $e^+e^-$ axis at angle $\theta_2$). The other relevant plane is determined by $\hat{\bm z}$ and the direction of the spin-1/2 hadron, $h_1$, [$\bm{P}_{h_1}= (P_{1T} \cos\phi_1,P_{1T}\sin\phi_1,P_{1L})$].
In this case, the transverse polarisation of $h_1$, in its helicity frame, as reached from the helicity frame of the fragmenting quark, is not directed along $\hat{\bm{n}}= (-\bm{P}_{h_2}\times \bm{P}_{h_1})/|\bm{P}_{h_2}\times \bm{P}_{h_1}|$ and has therefore to be projected out along this direction.
Moreover, two main and independent contributions appear: one still driven by the polarising FF, convoluted with the unpolarised TMD-FF for the hadron $h_2$, and another one driven by the Collins FF for the hadron $h_2$ and related to the production of a transversely polarised quark-antiquark pair. This last piece involves the two TMDs describing the fragmentation of a transversely polarised quark into a transversely polarised spin-1/2 hadron and manifests specific modulations in $\phi_1$.
The detailed calculation will be presented elsewhere~\cite{D'Alesio:2020xxx}. Here we give the final expression for the transverse polarisation along $\hat{\bm{n}}$, integrated over $\bm{P}_{1T}$ and adopting a Gaussian ansatz for the TMD-FFs. In particular we use:
\bea
D_{h/q}(z,p_{\perp}) & = & D_{h/q}(z)\, \frac{e^{-p_{\perp}^2/\langle p_{\perp}^2\rangle}}{\pi \langle p_{\perp}^2\rangle} \\
\Delta D_{h^\uparrow\!/q}(z,p_{\perp}) &= & \Delta D_{h^\uparrow\!/q}(z)  \frac{\sqrt{2e} \,p_{\perp}}{M_{\rm pol}} \frac{e^{-p_{\perp}^2/\langle p_{\perp}^2\rangle_{\rm pol}}}{\pi \langle p_{\perp}^2\rangle},
\label{Gaus}
\eea
where $M_{\rm pol}$ and $\langle p_\perp^2\rangle_{\rm pol}$ are related as follows
\be
\langle p_\perp^2\rangle_{\rm pol} = \frac{M_{\rm pol}^2}{M_{\rm pol}^2 + \langle p_{\perp}^2\rangle} \,\langle p_{\perp}^2\rangle\,.
\ee
By imposing $|\Delta D(z)|\le D(z)$ the positivity bound for the polarising FF, Eq.~(\ref{polFF}), is automatically fulfilled. At the same time, this form allows us to carry out analytically the integrations over transverse momenta (at ${\cal O}(p_\perp/(z\sqrt s))$. Notice that we use flavour independent Gaussian widths both for the unpolarised and the polarised FFs. The transverse polarisation can be finally expressed as \cite{D'Alesio:2020xxx}
\bea
&& {\cal P}_n(z_1,z_2) = \sqrt{\frac{e\pi}{2}}\frac{1}{M_{\rm pol}} \frac{\langle p_\perp^2\rangle_{\rm pol}^2}{\langle p_{\perp 1}^2\rangle}\nonumber\\
&& \mbox{} \times \frac{z_2}{\big\{[z_1(1-m_{h_1}^2/(z_1^2s))]^2\langle p_{\perp 2}^2\rangle +z_2^2 \langle p_\perp^2\rangle_{\rm pol}\big\}^{1/2}}\nonumber\\
&& \mbox{}\times \frac{\sum_{q} e^2_q (1 + \cos^2\theta_2)
\Delta D_{h_1^\uparrow/q}(z_1)D_{h_2/\bar q}(z_2)}{ \sum_{q}  e^2_q (1 + \cos^2\theta_2)
 D_{h_1/q}(z_1)D_{h_2/\bar q}(z_2)}\,.
\label{PolTh}
\eea
For its importance and later use we also give the first $p_\perp$-moment of the polarising function:
\bea
\Delta D_{h^\uparrow\!/q}^{(1)}(z) & = & \int\! d^2 \bm{p}_{\perp} \frac{p_{\perp}}{2 z m_h} \Delta D_{h^\uparrow\!/q}(z,p_{\perp}) =
D_{1T}^{\perp (1)}(z)\nonumber\\
&=& \sqrt{\frac{e}{2}}\frac{1}{z m_h} \frac{1}{M_{\rm pol}}\frac{\langle p^2_{\perp} \rangle^{2}_{\rm pol}}{\langle p^2_{\perp} \rangle}  \Delta D_{h^\uparrow\!/q }(z)\,,
\label{1mom}
\eea
where the last expression is obtained adopting the parametrization in Eq.~(\ref{Gaus}).
One can notice that ${\cal P}_n$, Eq.~(\ref{PolTh}), is directly sensitive to this quantity.

{\it Fit and Results.}
We can now proceed, using Eqs.~(\ref{PolTjet}) and (\ref{PolTh}), with the phenomenological analysis of Belle polarisation data for $\Lambda$ and $\bar\Lambda$ production, measured at $\sqrt s = 10.58$ GeV~\cite{Guan:2018ckx}. Two sets are available: one for inclusive production as a function of $p_{\perp}$ (the $\Lambda$ transverse momentum w.r.t.~the thrust axis), for different energy fractions, $z_\Lambda$, (32 data points) and a second one for the associated production of $\Lambda$ with light hadrons, namely charged pions and kaons, as a function of the energy fractions $z_\Lambda$ and $z_\pi(z_K)$ (128 data points). Notice that here we consider the transverse polarisation for inclusive $\Lambda$ particles, namely those directly produced from $q\bar q$ fragmentation and those indirectly produced from strong decays.

We parameterize the $z$-dependent part (adopting the light-cone momentum fraction) $\Delta D_{\Lambda^\uparrow\!/q}(z)$ as:
\be
\Delta D_{\Lambda^\uparrow\!/q}(z) = N_q z^{a_q} (1-z)^{b_q} \frac{(a_q+b_q)^{(a_q+b_q)}}{a_q^{a_q}b_q^{b_q}} D_{\Lambda/q}(z)\,,
\label{Deltaz}
\ee
where $|N_q|\le 1$ and $q=u,d,s, \rm{sea}$ (see below). This guaranties $|\Delta D(z)|\le D(z)$.

For the unpolarised FFs we adopt the DSS07 set~\cite{deFlorian:2007aj}, for pions and kaons, and the AKK08 set~\cite{Albino:2008fy} for $\Lambda$'s. Since all data are at fixed energy scale no evolution is implied in this extraction. For the unpolarised Gaussian widths we use $\langle p^2_{\perp} \rangle = 0.2$ GeV$^2$, as extracted in Ref.~\cite{Anselmino:2005nn}, both for light and heavy hadrons. We have checked that varying this value has a little effect in the final results.
Concerning the $\Lambda$ FF set, all available parameterizations are given for $\Lambda+\bar\Lambda$, including the AKK08 set, which adopts $z_p$ as scaling variable. We then separate the two contributions as follows:
\bea
D_{\Lambda/q}(z_p) &=& D^{\Lambda+\bar\Lambda}_q(z_p)\,\frac{1}{1+(1-z_p)^s}\\
D_{\bar\Lambda/q}(z_p) &=& D^{\Lambda+\bar\Lambda}_q(z_p)\, \frac{(1-z_p)^s}{1+(1-z_p)^s}\,,
\eea
where the power $s$ has been set to 1 (we checked that higher values have a low impact on the present study). Notice that all transformations among the different scaling variables ($z,z_h,z_p$) involved, Eqs.~(\ref{zhp}), (\ref{zpzh}), have been taken properly into account.

In order to access the $p_\perp$ dependence of the polarising FF, data in the thrust-plane frame would be ideal. On the other hand, the experimental accuracy in extracting them, requiring the reconstruction of the thrust axis, is more problematic. This reflects on the quality of the fit. The analysis of associated production data, extremely powerful in accessing flavour separation, is somehow easier from the experimental point of view, but phenomenologically more complex and model dependent.
We have therefore tried to perform a global fit including both data sets, but paying attention to data at very large $z_h$. In particular, for the inclusive production data set, which, as already pointed out, presents some difficulties, we cut out the largest $z_\Lambda$ bin, extending from 0.5 to 0.9.
For the associated production we adopt a similar choice, excluding data where the energy fractions for both hadrons are too large. We have then imposed the following cuts: $z_\Lambda \le 0.5$ for the inclusive production and $z_{\pi,K} \le 0.5$ for the associated production data set. This leaves us with 24 + 96 = 120 data points, still allowing to probe, at least in the $\Lambda h$ data set, large values of $z_\Lambda$. We notice that relaxing the cut on $z_\pi$, the quality of the fit would not change.

Concerning the parameters of the $z$-dependent part, Eq.~(\ref{Deltaz}), the best global fit is obtained adopting the following set:
\be
N_u, \; N_d, \; N_s, \; N_{\rm sea}, \; a_s, \; b_u, \; b_{\rm sea}\,,
\ee
with all other $a$ and $b$ parameters set to zero. This means that,
with $\langle p^2_{\perp} \rangle_{\rm pol}$ (Eq.~(\ref{Gaus})), we have 8 free parameters.

We have to mention that \textit{simpler} fits with only two polarising FFs, for $u(=d)$ and $s$ quarks, or even those without any sea contribution give much higher $\chi^2_{\rm dof}$'s. The same happens if no appropriate modulation in $z$ is included. See comments below.

\begin{table}[b!]
\caption{
Best values of the 8 free parameters fixing the polarising FF (Eqs.~(\ref{Gaus}), (\ref{Deltaz})) for $u,d,s$ and sea quarks, as obtained by fitting Belle data~\cite{Guan:2018ckx}. The statistical errors quoted for each parameter correspond to the shaded uncertainty areas in Figs.~\ref{fig:Lj} and \ref{fig:Lh}, as explained in the text and in the Appendix of Ref.~\cite{Anselmino:2008sga}.
\label{fitpar}}
\vskip 18pt
\renewcommand{\tabcolsep}{0.4pc} 
\renewcommand{\arraystretch}{1.2} 
\begin{tabular}{@{}ll}
 \hline
 $N_{u} = 0.47^{+0.32}_{-0.20}$ & $N_{d} =  -0.32^{+0.13}_{-0.13}$ \\
 $N_{s} = -0.57^{+0.29}_{-0.43}$ & $N_{\rm sea} =  -0.27^{+0.12}_{-0.20}$\\
 $a_s = 2.30^{+1.08}_{-0.91}$ &\\
 $b_u = 3.50^{+2.33}_{-1.82}$ & $b_{\rm sea}  = 2.60^{+2.60}_{-1.74}$ \\
 \hline
  $\langle p^2_{\perp} \rangle_{\rm pol} = 0.10^{+0.02}_{-0.02}$ GeV$^2$ & \\
 \hline
\end{tabular}
\end{table}

\begin{figure}[!t]
\includegraphics[trim =  0 50 0 100,clip,width=8.5cm]{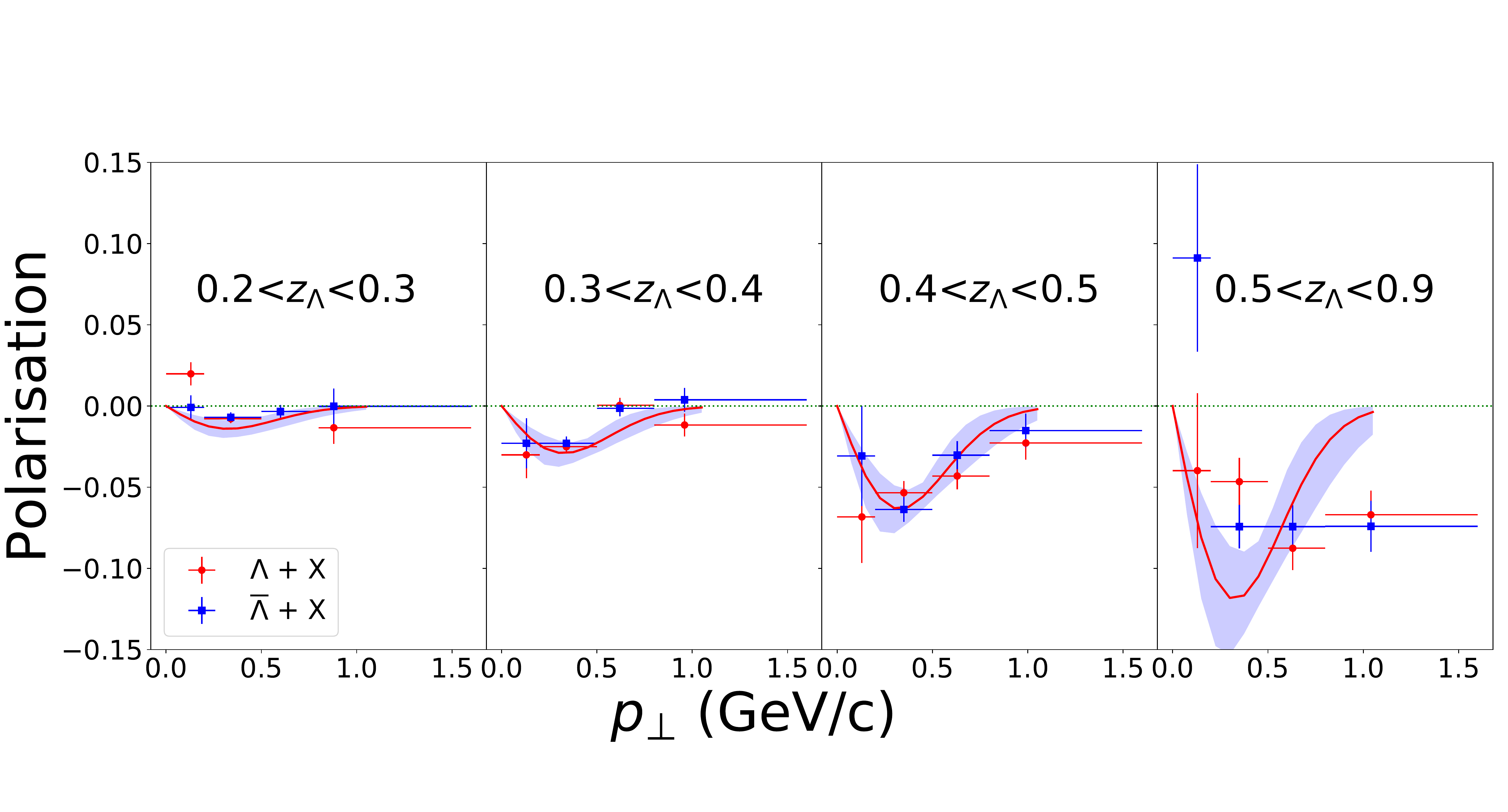}
\caption{Best-fit estimates of the transverse polarisation for inclusive $\Lambda$ and $\bar\Lambda$ production in $e^+e^-\to \Lambda(\rm{jet}) + X$ (thrust-plane frame) as a function of $p_\perp$ for different $z_\Lambda$ bins (energy fractions), compared against Belle data~\cite{Guan:2018ckx}. The statistical uncertainty bands, at 2$\sigma$ level, are also shown. Notice that curves for $\Lambda$ and $\bar\Lambda$ coincide and that data in the rightmost panel are not included in the fit.}
\label{fig:Lj}
\end{figure}

\begin{figure}[!t]
\includegraphics[trim =  0 50 0 100,clip,width=8.5cm]{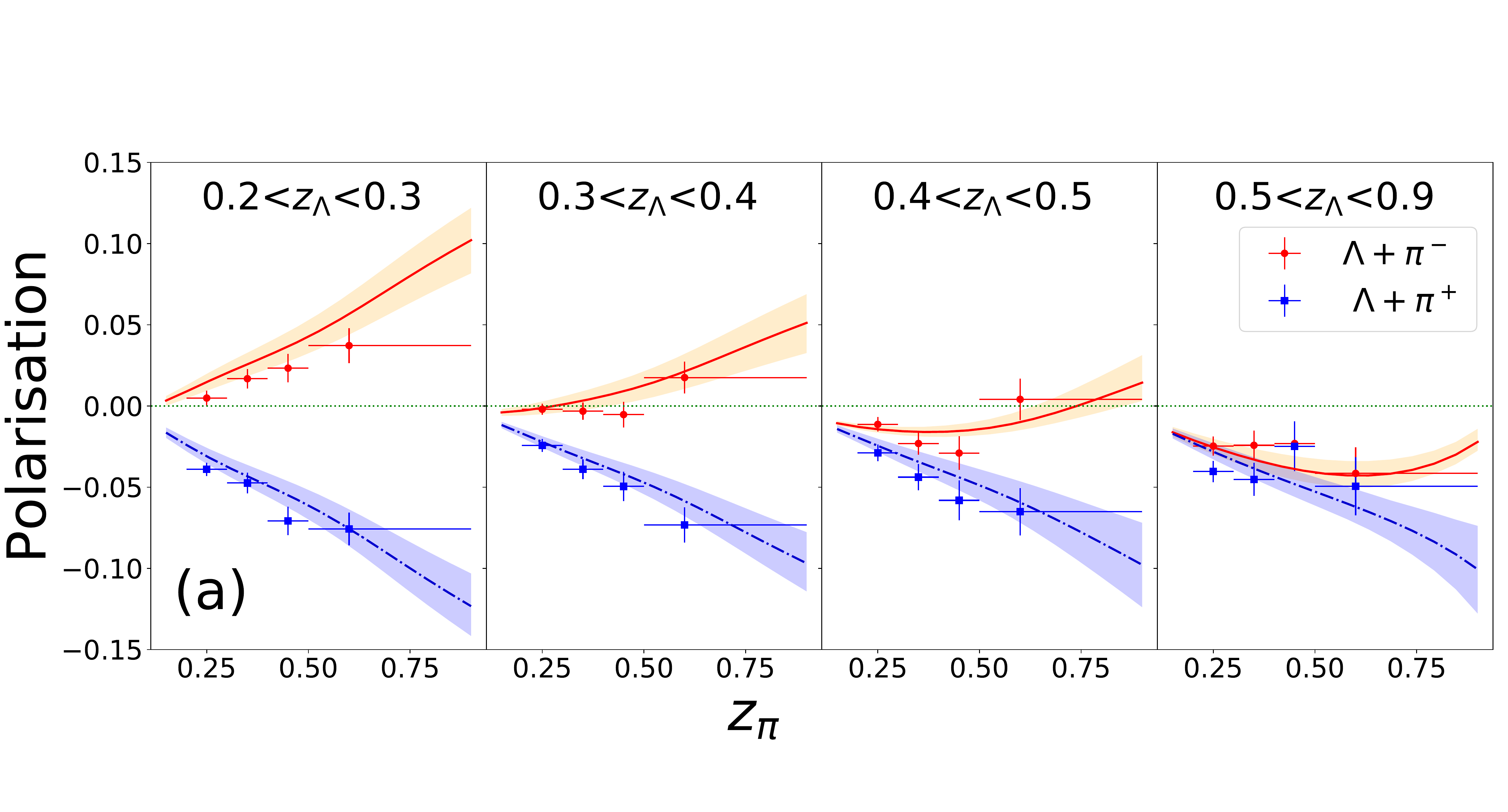}
\includegraphics[trim =  0 50 0 100,clip,width=8.5cm]{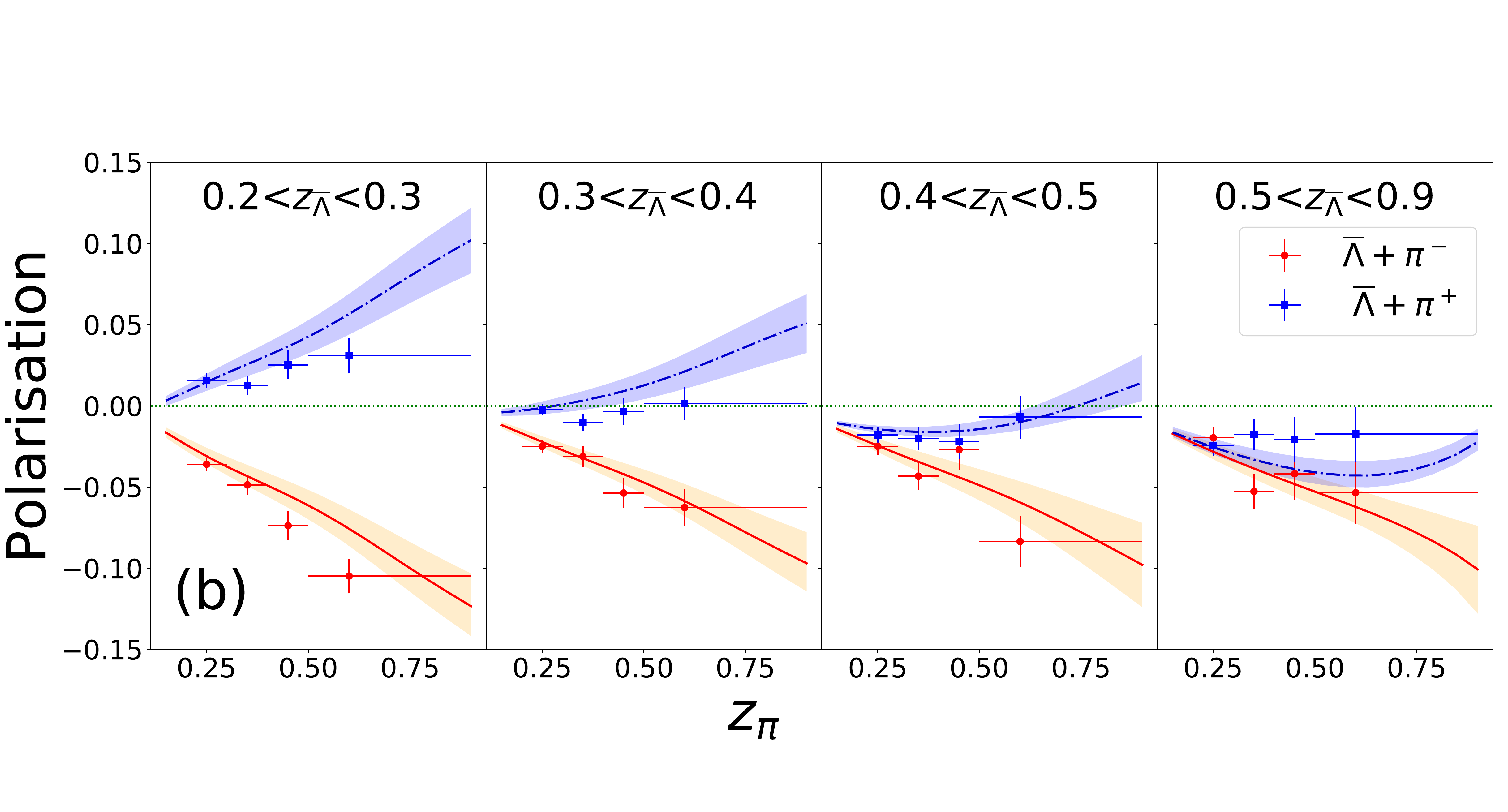}
\includegraphics[trim =  0 50 0 100,clip,width=8.5cm]{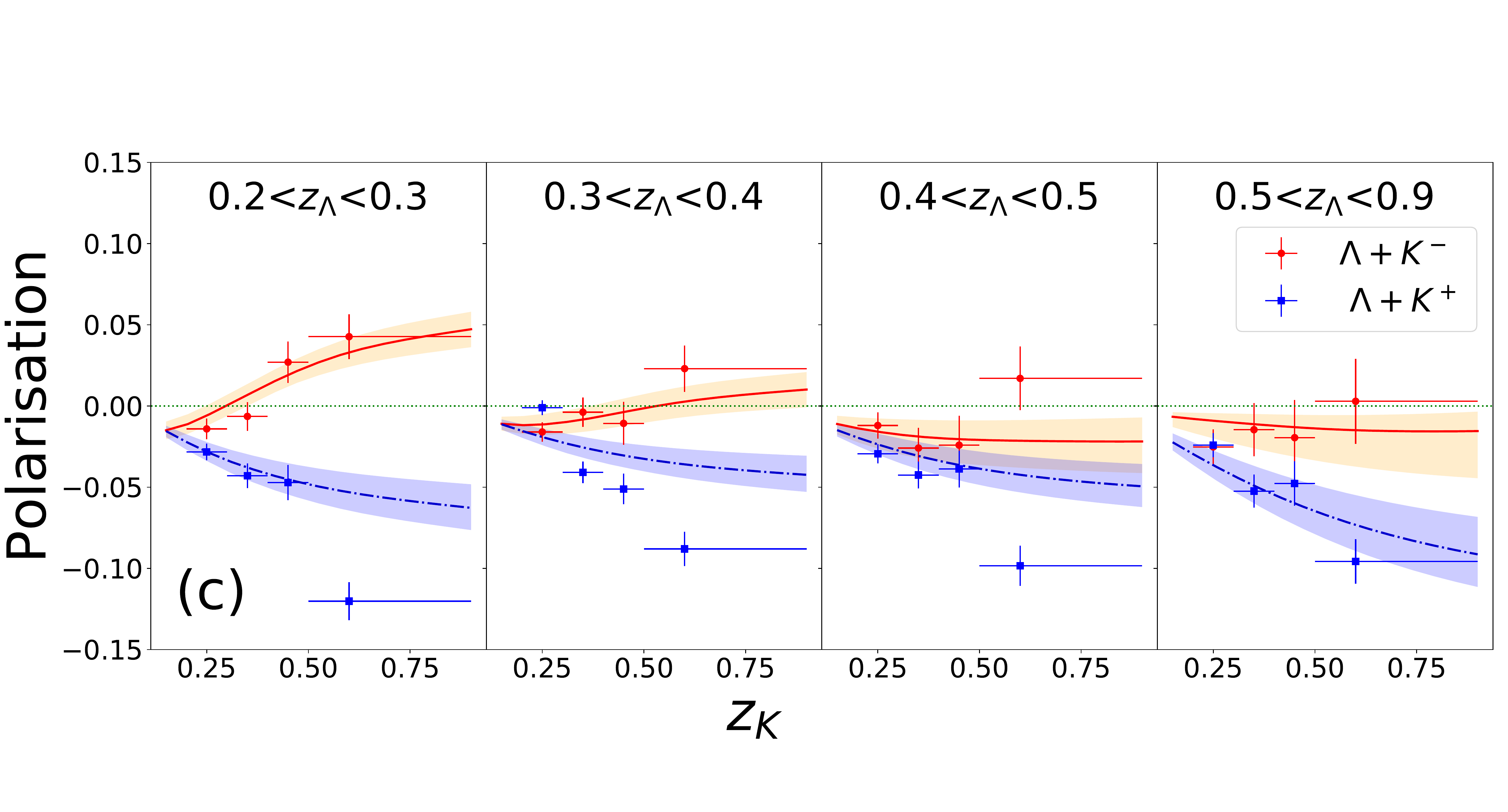}
\includegraphics[trim =  0 50 0 100,clip,width=8.5cm]{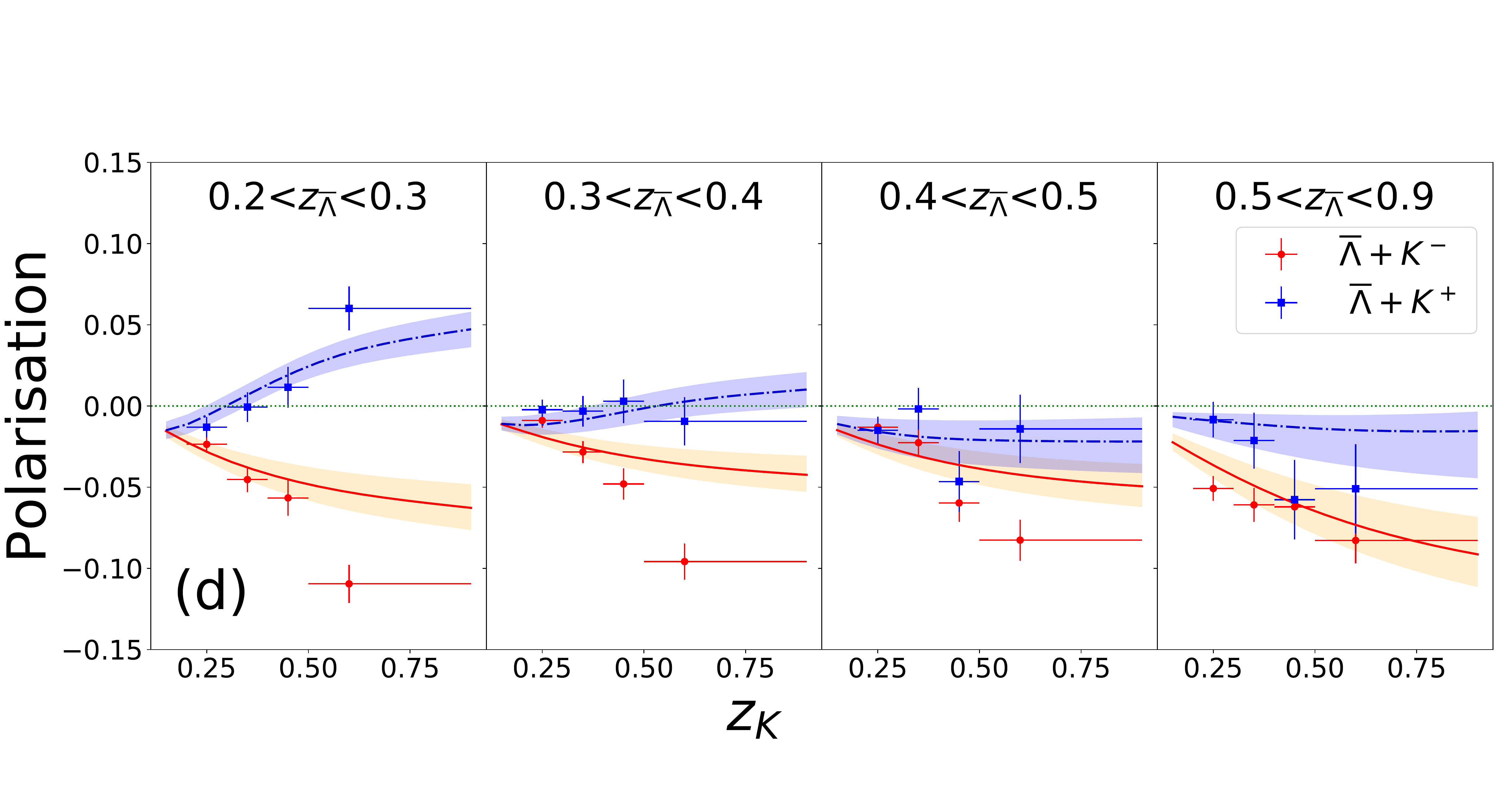}
\caption{Best-fit estimates of the transverse polarisation for $\Lambda$ and $\bar\Lambda$ production in $e^+e^-\to \Lambda(\bar\Lambda) h + X$, for $\Lambda\pi^\pm$ (a), $\bar\Lambda\pi^\pm$ (b), $\Lambda K^\pm$ (c), $\bar\Lambda K^\pm$ (d), as a function of $z_{h}$ (of the associated hadron) for different $z_\Lambda$ bins. Data are from Belle~\cite{Guan:2018ckx}. The statistical uncertainty bands, at 2$\sigma$ level, are also shown. Notice that data for $z_{\pi,K}>0.5$ are not included in the fit.}
\label{fig:Lh}
\end{figure}

Table~\ref{fitpar} reports the values of the parameters as determined by the best-fitting procedure. The corresponding results, compared to Belle data~\cite{Guan:2018ckx}, are shown in Figs.~\ref{fig:Lj} and \ref{fig:Lh}, respectively for the inclusive and the associated $\Lambda$ hadron ($\pi^\pm$ and $K^\pm$) production. The quality of the fit is reasonably good with a $\chi^2_{\rm dof} = 1.94$ and with $\chi^2_{\rm points} = 2.75, 1.55, 1.61$ for jet, pion and kaon data subsets. We notice that a fit restricted only to associated production data would give a $\chi^2_{\rm dof} =1.26$ with $\chi^2_{\rm points}= 0.8, 1.5$ for pion and kaon data subsets (we will come back to this).
The shaded areas, corresponding to a 2$\sigma$ uncertainty, are computed according to the procedure explained in the Appendix of Ref.~\cite{Anselmino:2008sga}. More precisely, we have allowed the set of best fit parameters to vary in such a way that the corresponding new curves have a total $\chi^2\le \chi^2_{\rm min} + \Delta\chi^2$.
All these new curves lie inside the shaded area. The chosen value of $\Delta\chi^2= 15.79$, for our eight-parameter fit, is such that the probability to find the ‘‘true’’ result inside the shaded band is 95.45\%. The quoted statistical errors in Table~\ref{fitpar}
correspond to these shaded areas.

\begin{figure}[!thb]
\includegraphics[trim =  150 0 300 0,clip,width=4.2cm]{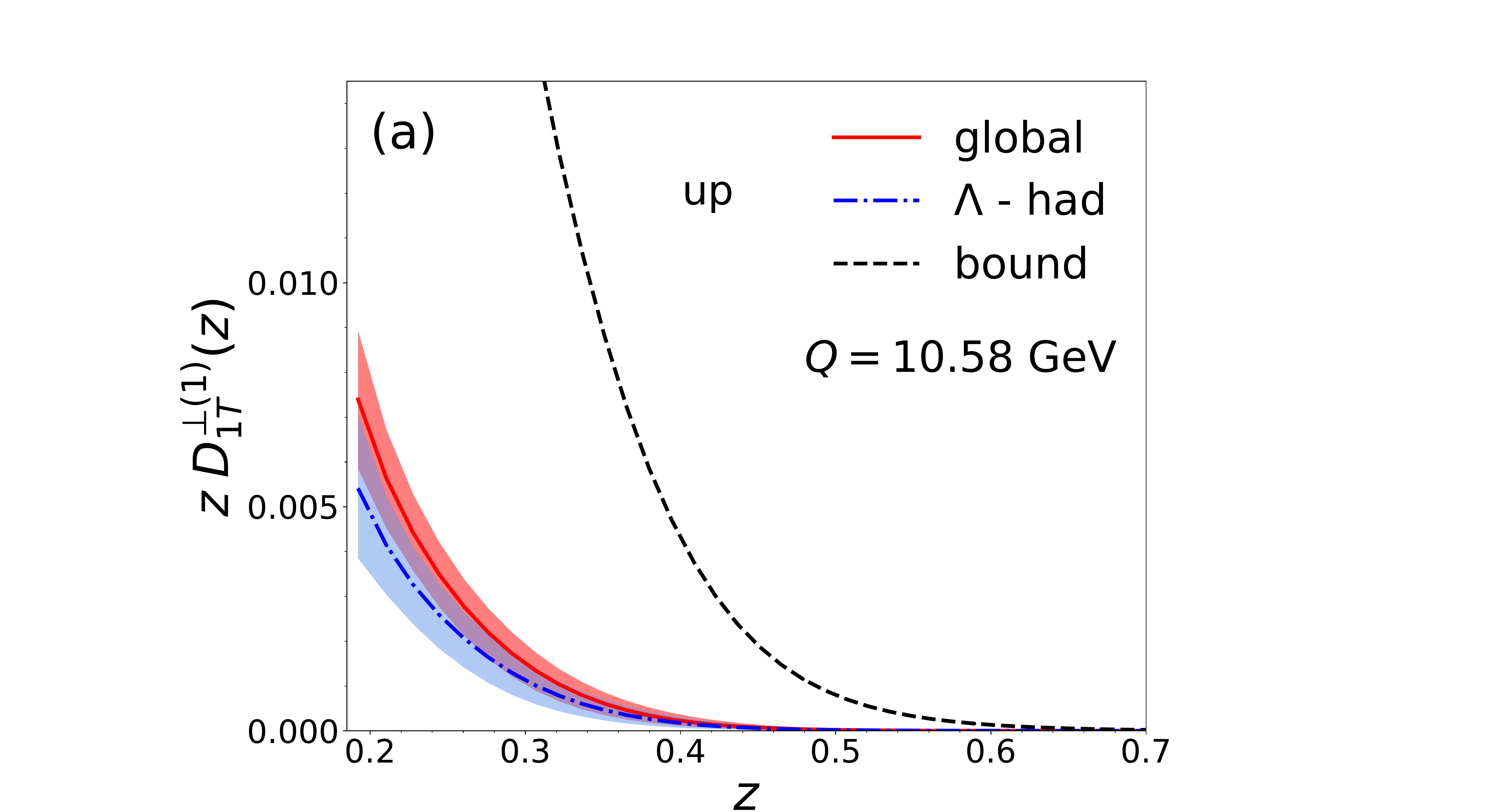}
\includegraphics[trim =  150 0 300 0,clip,width=4.2cm]{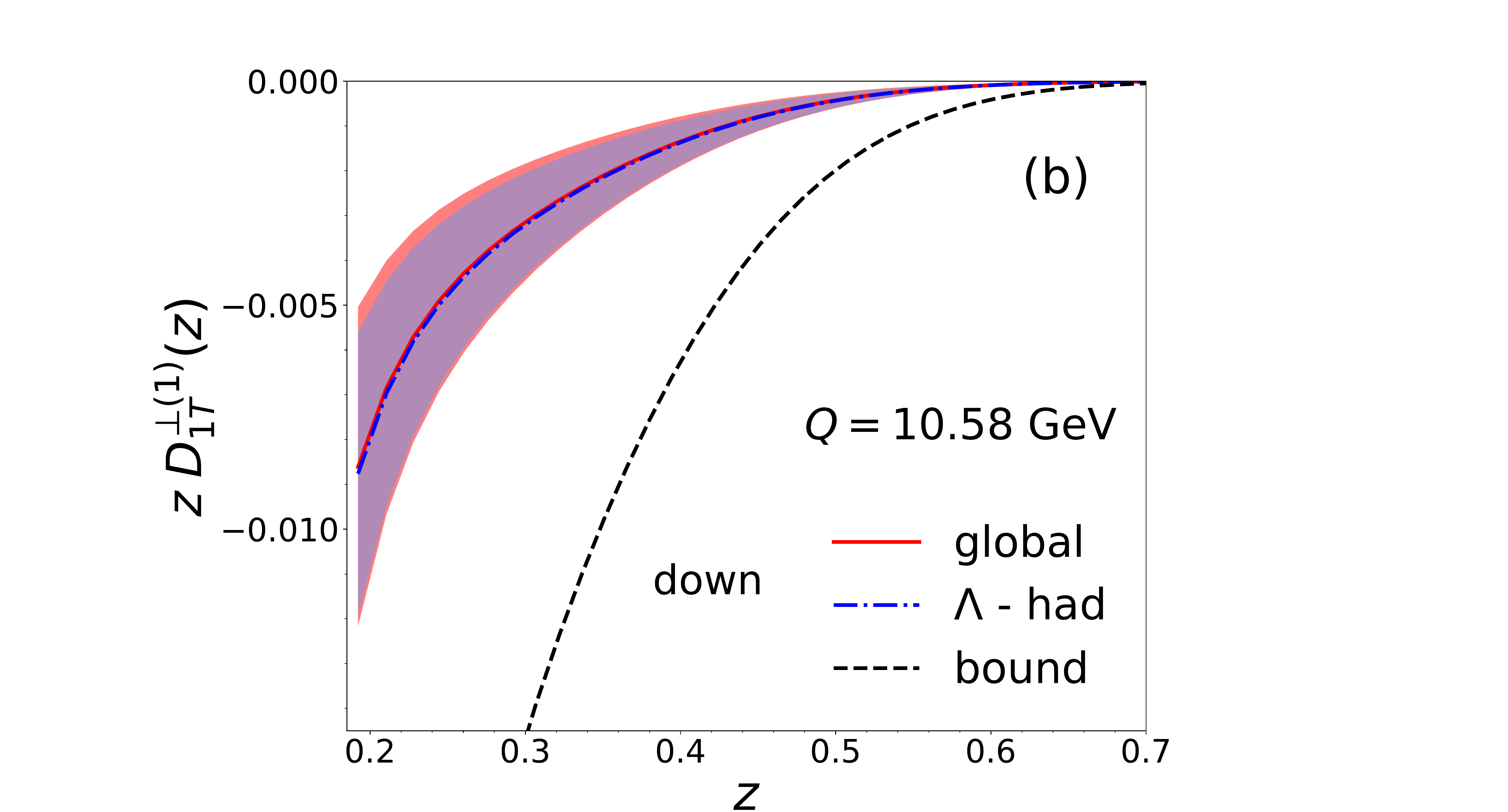}

\includegraphics[trim =  150 0 300 0,clip,width=4.2cm]{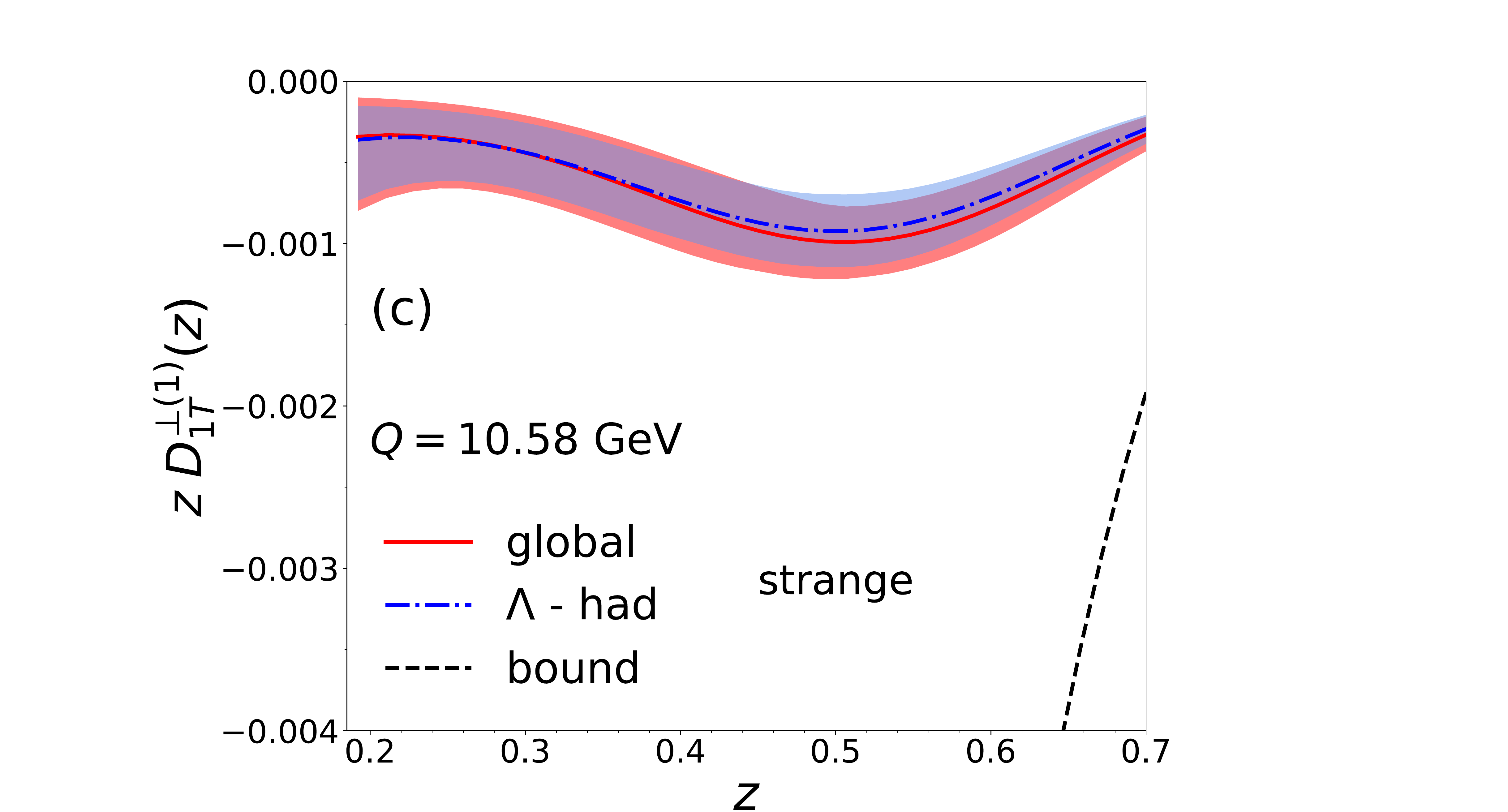}
\includegraphics[trim =  150 0 300 0,clip,width=4.2cm]{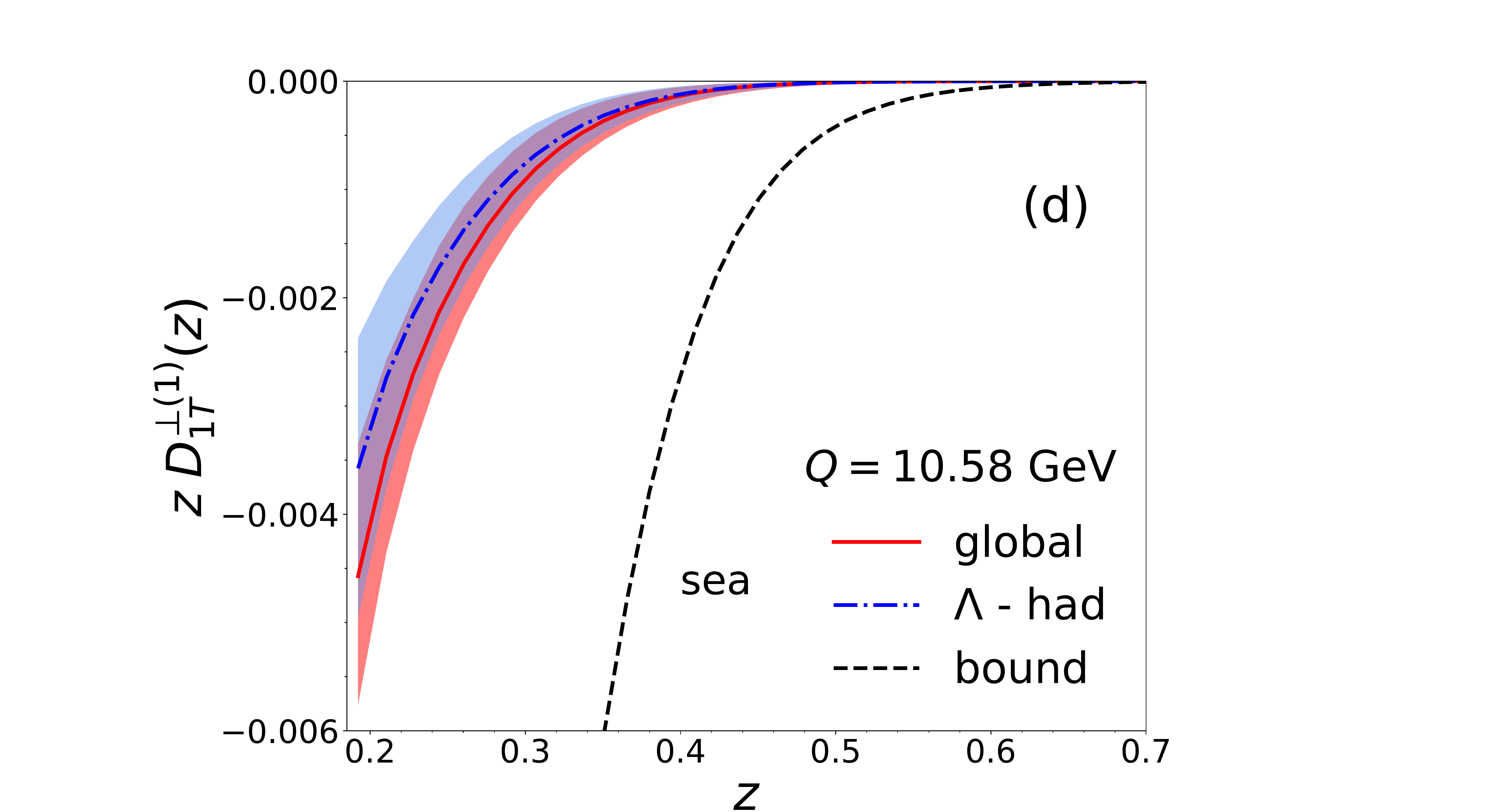}
\caption{First moments of the polarising fragmentation functions, see Eq.~(\ref{1mom}), for the up (a), down (b), strange (c) and sea (d) quarks, as obtained from the global fit (red solid lines) and the $\Lambda$-hadron fit (blue dot-dashed lines). The corresponding statistical uncertainty bands (at 2$\sigma$ level), as well as the positivity bounds (black dashed lines), are also shown.}
\label{fig:1stm}
\end{figure}

\begin{figure}[bht]
\includegraphics[trim =  200 0 200 50,clip,width=6.cm]{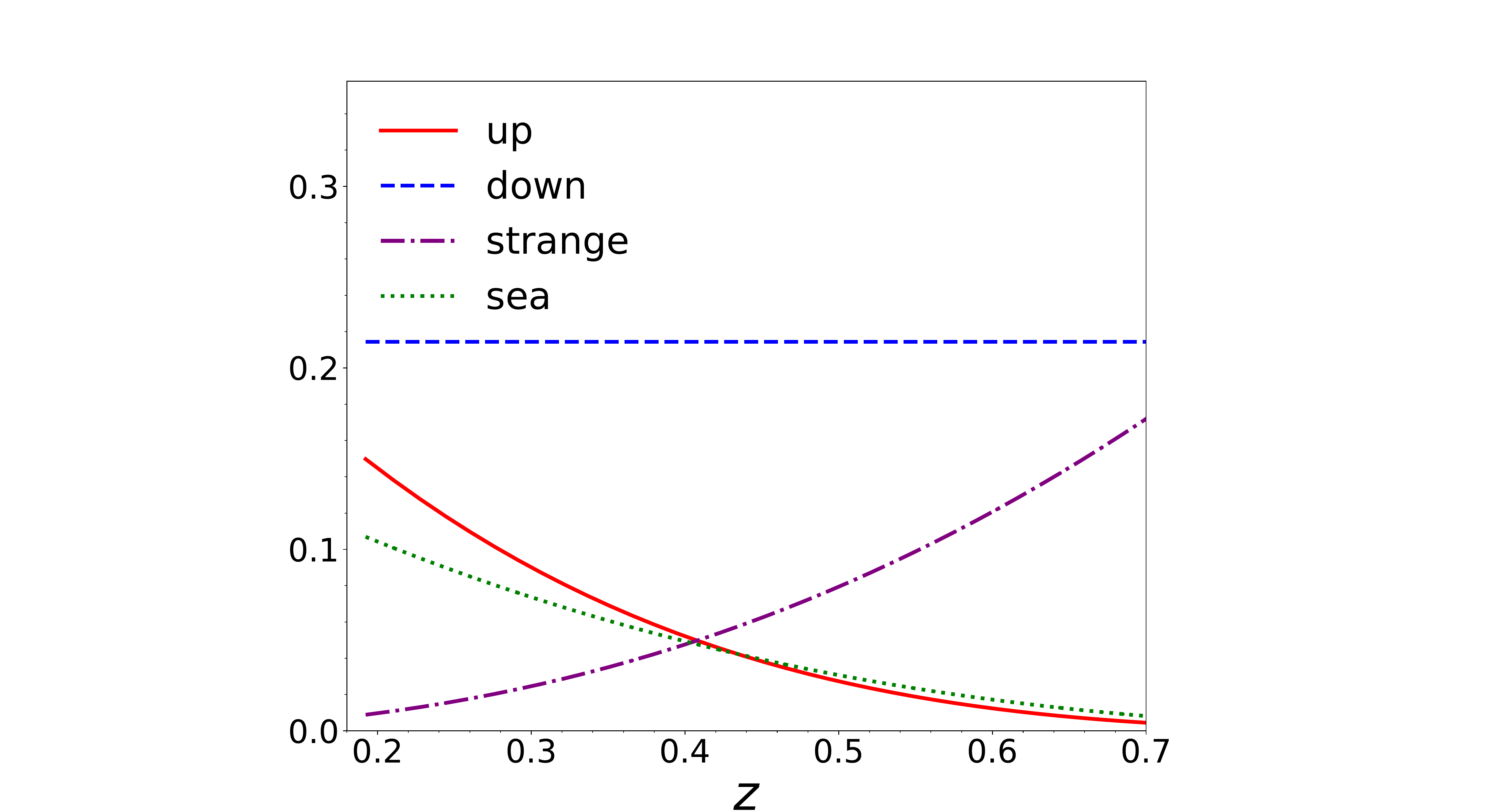}
\caption{Ratios of the absolute values of the first moments of the polarising FFs with respect to their corresponding positivity bounds for the up (red solid line), down (blue dashed line), strange (purple dot-dashed line) and sea (green dotted line) quarks, as obtained from the global fit.}
\label{fig:ratios}
\end{figure}

As mentioned above, a fit limiting only to the associated production data gives a much better result. Even though the resulting best-fit parameters are a bit different, the corresponding first moment, Eq.~(\ref{1mom}), is quite stable and the two extractions lead to consistent results. This is shown in Fig.~\ref{fig:1stm}, where we present the first moments as obtained in the global fit (red solid lines) and the corresponding ones obtained by fitting only the associated production data (blue dot-dashed lines). As one can see, they are well consistent within their uncertainty bands, and in some cases (down and strange quarks) almost indistinguishable, supporting the reliability of this extraction. Moreover, they lay within their positivity bounds (black dotted lines). In Fig.~\ref{fig:ratios}, to better realise their sizes, as well as their behaviour, we show, for the global fit, the ratios of the absolute value of the first moments w.r.t.~the corresponding positivity bounds.

Some comments are in order.
For the inclusive production case, the description is clearly less good (the relative $\chi^2_{\rm points}$ being around 2.8). On the other hand, one would expect ${\cal P}_T=0$ at $p_\perp=0$, as well as ${\cal P}_T(\bar\Lambda)={\cal P}_T(\Lambda)$, a feature not clearly visible in the data (see Fig.~\ref{fig:Lj}). This somehow increases the tension with the other data set, reducing the quality of the global fit.

Moving to the associated production data set we start observing that charge-conjugation symmetry implies ${\cal P}_n(\Lambda h^+)= {\cal P}_n(\bar\Lambda h^-)$ and this is what happens for our estimates; in this respect also the data are quite consistent (Fig.~\ref{fig:Lh}).
It is definitely illuminating to consider $\Lambda\pi$ and $\Lambda K$ data separately: for medium $z_\pi$ ($z_K$), where the corresponding favoured unpolarised FFs dominate, we can say that $\Lambda \pi^-$,
$\Lambda \pi^+$, $\Lambda K^+$ data help in constraining the sign and the size of the polarising FFs respectively for up, down and strange quarks.
More precisely, the relative sign between the polarising FF for up ($N_u>0$) and down ($N_d<0$) quarks can be traced back to the opposite sign between ${\cal P}_n(\Lambda \pi^-)$ ($>0$) and ${\cal P}_n(\Lambda \pi^+)$ ($<0$).
This motivates and explains the use of a different polarising FF for these two flavours (Figs.~\ref{fig:1stm}a, ~\ref{fig:1stm}b), even if the corresponding unpolarised FFs are equal. Moreover, the reduction in size of ${\cal P}_n(\Lambda \pi^-)$ w.r.t.~${\cal P}_n(\Lambda \pi^+)$, reaching negative values for $z_\Lambda\ge 0.4$, requires a larger suppression of the up polarising FF w.r.t.~the down one at large $z$ (resulting in a large $b_u$, with $b_d=0$, see also Fig.~\ref{fig:ratios}, red solid and blue dashed lines). For $z_\Lambda \le 0.4$ the sea quarks start playing some role, becoming important at $z_\Lambda \le 0.3$. For instance, for ${\cal P}_n(\Lambda \pi^+)$, where the up and down contributions almost cancel between each other for these $z_\Lambda$ values, it is the negative sea polarising FF that leads to large, and negative, values of the transverse polarisation. Similarly, in ${\cal P}_n(\Lambda \pi^-)$, still for $z_\Lambda \le 0.3$, this is responsible for the partial reduction of the very large piece driven by the up polarising FF, coupled to the favoured unpolarised $\pi^-$ FF and weighted by a large relative charge factor. We can then understand the negative sign of the sea polarising FF (Fig.~\ref{fig:1stm}d) and its strong suppression at large $z$ (Fig.~\ref{fig:ratios}, green dotted line).

The description of $\Lambda$-kaon data follows a similar pattern, with some peculiarities.
We can easily understand the negative values of ${\cal P}_n(\Lambda K^+$) at medium $z_\Lambda$, being driven by a sizeable and negative $\Delta D_{\Lambda^\uparrow\!/s}$ (Fig.~\ref{fig:1stm}c), coupled to the leading FF $D_{K^+/\!\bar s}$. When moving to smaller $z_\Lambda$, this contribution is suppressed (due to the large $a_s$ value, see also Fig.~\ref{fig:ratios}, purple dot-dashed line) and once again it is the negative sea quark polarising FF, $\Delta D_{\Lambda^\uparrow\!/\bar u}$, which lead to large and negative ${\cal P}_n(\Lambda K^+$) values. For ${\cal P}_n(\Lambda K^-$) at medium-large $z_\Lambda$ all contributions are negligible: because of the suppression at large $z$ of the corresponding polarising FF (up and sea quarks), or because they are coupled to sub-leading sea unpolarised $K^-$ FFs. On the other hand, at small $z_\Lambda$ the up quark dominates, leading to large and positive ${\cal P}_n$, slightly reduced by the negative sea polarising FF, $\Delta D_{\Lambda^\uparrow\!/\bar s}$, coupled to the leading unpolarised FF $D_{K^-/\!s}$. Notice that the relative size of the valence kaon unpolarised FF for up quark w.r.t.~the strange one plays a crucial role.

Of course, all the above reasoning can be applied to the corresponding $\bar\Lambda h$ data sets.

All these findings are in perfect agreement with the qualitative expectations discussed in Ref.~\cite{Guan:2018ckx}, with extra information on the down polarising FF and supported by a quantitative extraction.

{\it Conclusions.}
The recent data from Belle Collaboration for the transverse $\Lambda/\bar\Lambda$ polarisation have been used to extract, for the first time, the TMD polarising fragmentation function of $\Lambda$ hyperons. A quite good separation in flavours has been achieved, thanks to the experimental results for associated production in conjunction with pion and kaons.
From this analysis it turns out the need for different polarising FFs for the three valence contributions.
The relative sign, as well as the size, of the favoured polarising FFs for up, down and strange quarks, have been extracted with good accuracy and are well under control. Similarly, the need of a sea quark polarising FF emerges quite clearly.
Concerning their $p_\perp$ dependence, one has to keep in mind that the corresponding one for the unpolarised FF is still unknown, and can only be guessed; moreover, data for the inclusive production, relevant in this respect, manifests some tension w.r.t.~the associated production data set. Nonetheless, we have been able to extract, within a Gaussian factorised ansatz, a reasonable information on it.
New data with higher statistics will be extremely useful, and complementary studies in other processes will certainly help towards a deeper understanding of this important TMD fragmentation function as well as of the observed spontaneous transverse hyperon polarisation.

{\it Acknowledgement.}
We thank M.~Anselmino for useful discussions.
This work is supported by the European Union's Horizon 2020 research and innovation programme under grant agreement No.~824093 (STRONG2020).


\end{document}